# All-atom *ab* initio folding of a diverse set of proteins

Jae Shick Yang[1], William W. Chen[2,1], Jeffrey Skolnick[3], and Eugene I. Shakhnovich[1,*]

[1]Department of Chemistry and Chemical Biology

[2]Department of Biophysics

Harvard University

Cambridge, Massachusetts 02138

[3]Center of Excellence in Bioinformatics

University at Buffalo

Buffalo, New York 14203

Correspondence: e-mail: eugene@belok.harvard.edu, phone: 617-495-4130, fax: 617-384-9228.

## Abstract

Natural proteins fold to a unique, thermodynamically dominant state. Modeling of the folding process and prediction of the native fold of proteins are two major unsolved problems in biophysics. Here, we show successful all-atom *ab initio* folding of a representative diverse set of proteins, using a minimalist transferable energy model that consists of two-body atom-atom interactions, hydrogen-bonding, and a local sequence energy term that models sequence-specific chain stiffness. Starting from a random coil, the native-like structure was observed during replica exchange Monte Carlo (REMC) simulation for most proteins regardless of their structural classes; the lowest energy structure was close to native- in the range of 2-6 Å root-mean-square deviation (RMSD). Our results demonstrate that the successful folding of a protein chain to its native state is governed by only a few crucial energetic terms.

## Introduction

*Ab initio* simulations have been used to study folding kinetics/dynamics and to predict protein structure (Duan and Kollman, 1998; Klimov and Thirumalai, 2000; Lazaridis and Karplus, 1997; Liwo et al., 2005; Snow et al., 2002). During all-atom molecular dynamics simulation with explicit water of a small protein, villin headpiece, Duan and Kollman (Duan and Kollman, 1998) observed a marginally stable conformation which resembles the native structure of the protein. In the reduced representation models, Liwo *et al.* (Liwo et al., 2005) optimized the potential for seven proteins with various topologies and tried to fold the same proteins. They could observe the native-like structure for 5/7 proteins during Langevin dynamics simulations.

Recently, we proposed a method for deriving an all-atom protein folding potential, the μ-potential (Hubner et al., 2005; Kussell et al., 2002). The potential relies solely on an atom-typing scheme and pairwise interaction scoring terms for each residue pair. This potential was derived for and tested on a three-helix bundle protein, Protein Data Bank (PDB) ID code 1BDD (Kussell et al., 2002). Using this formulation, starting from a random state, we folded 1BDD to less than a 2 Å $C_\alpha$ RMSD from native. Here, we describe work on combining the transferable μ-potential with an improved hydrogen-bond potential as well as a term describing local conformational propensities that can fold various classes of proteins in an all-atom representation.

## Results

The atomic model uses the energy function of Eq. (1) (See Experimental Procedures). It contains only three adjustable parameters corresponding to the relative weights of energy contributions and the weight factor for β strand hydrogen-bonding. The three parameters *a, b, c* were derived on 5 out of 18 proteins used as the training set and transferred without further changes to the remaining 13 proteins used as the testing set (see Experimental Procedures). Native-like structures are observed in the course of REMC folding simulations for most proteins regardless of their class (Table 1). However, generation of such structures is not sufficient- an objective criterion for structure selection that does not rely on the known native structure needs to be employed. Using the minimum energy structure as the selection criterion, the all-atom simulations fold most proteins to moderate resolution with native-like packing and topology, and in some cases to high resolution (e.g. 1ENH and 1GJS). Although some proteins are predicted with relatively high RMSD to the native structure, they have the correct topology, as discussed in the followings.

**Folding Proteins in Training and Test Sets**

For proteins from the training set, we successfully folded the two WW domain proteins (1E0L and 1I6C), which are three-stranded β-structures. The predicted structure of 1I6C matches the experimental structure within an RMSD of 2.7 Å (Figure 1A). For 1E0G, a 48-residue α+β protein consisting of two sheets and three helices, the predicted structure has the correct topology except for a missing helix in the middle of the chain, which is marked by arrows (Figure 1B). The predicted topology for a 48-residue three-helix bundle protein, 1ENH shows excellent agreement with the native structure (Figure 1C).

Most importantly, successful results were observed for many proteins in the test set (Figure 2). 1IGD, a 57-residue α/β protein composed of one helix and two hairpins that form a four member β-sheet, is folded to the correct topology (Figure 2A). Moreover, the final structure is very close to the native structure over the entire length of the protein. The only difference between the predicted and experimental structures is a shift in the relative orientation of the middle helix against the strands. 1CLB, a 75 residue $\alpha+\beta$ protein consisting of four helices and two strands, was also folded correctly (Figure 2B). Our model correctly generated all helices in this protein, missing only two short strands (marked by arrows). Two 3-helix bundle proteins, 1GJS and 1LQ7 were also correctly folded (Figure 2C, D).

**The Energy Landscape**

Since the validity of this procedure depends crucially on the ability of the potential in the detailed atomistic model to identify near-native states as lowest energy conformations, we next examined the nature of the energy landscape presented as energy-RMSD scatter plot (Figure 3). An important issue is whether deviations from the native conformations for the lowest energy structure found in folding simulations are due to flaws in the potential or insufficient sampling that prevented us from reaching the lowest energy conformations. To address this issue we ran control simulations following the same protocol but starting from the native conformations (shown as blue data points on Figure 3). It can be seen that the lowest energy conformations have quite low RMSD and that generally simulations starting from native conformations found conformations with lower energy and lower RMSD than runs that started from the fully unfolded chain, especially in cases when folding simulations resulted in apparently lowest energy conformations with high RMSD as in the case of a longer protein 1CSP (Figure 3D). This observation suggests that insufficient sampling is mostly responsible for most significant deviations from the native conformations in folding runs. At low backbone RMSD, conformations span a broad range of energies as can be seen in Figure 3. A detailed analysis of these low RMSD conformations showed that such spread of energies is due mostly to variations in side-chain packing. The appearance of discrete "stripes" in the energy landscape is due to mirror image topologies for helical bundles that sometimes may have pretty low energy (Figure 3C). Nevertheless our potential and *ab initio* simulations are able to distinguish the native fold and the mirror image.

**Native-like topologies of High-RMSD Structures**

While *ab initio* simulations predicted native-like structures at modest resolution for most proteins, for some proteins they generated topologically correct yet high RMSD models. In the protein 1BA5, a 46-residue α protein, three helices are packed in a mirror-image like manner, and the loop between helices marked by an arrow is too short when compared to the experimental structure (Figure 4A). One strand in the C terminus is not correctly formed in the case of 1SHFA, which prevents the chain from forming the closed β barrel structure (Figure 4B). For 1CSP, two strands in the N terminus have different alignment from the native, resulting in high RMSD value (Figure 4C). In both cases of 1SHFA and 1CSP, the control runs starting from the native state show lower RMSD and energy than runs starting from random coils and the gap in energies between runs starting from native and random coil structures is much larger than that for other proteins (Table 1). This means that the failure for these two proteins to reach a more native-like state may be due to conformational sampling. These two proteins are β-rich ones with complex topology. They fold more slowly than α proteins, making it computationally more difficult to find the equilibrium state within the simulation time. For the 70-residue 1AIL, the two helices are correctly predicted while the third helix is not formed properly (Figure 4D). Overall, although the predicted structures for these proteins have relatively high RMSD, they are far from being completely wrong, as can be seen from their *MaxSub* values (Siew et al., 2000) which are comparable to those of proteins folded at modest resolution, e.g. 1LQ7 (Table 1).

## Discussion

### The Energy Model

The potential energy function presented here consists of three terms. While it is not easy to pinpoint exactly the contributions of each of these terms, it is possible to show that the proteins are not folding for a trivial reason due to any one of the energy terms. To that end extensive controls were carried out in our previous publication that dealt with a simplified version of the potential (Hubner et al., 2005). Here we addressed this issue for the full potentials used in this work. For example, one possibility can be that the knowledge-based local sequence energetic term simply biases the secondary structure, and non-specific collapse of these secondary structure elements leads to the formation of a folded protein. In order to show that the success of simulations is not simply due to correct assignment of secondary structure and chain collapse, we performed a control simulation in which the contact potential was randomly reshuffled. In the case of the easily folded protein 1ENH, the end result with the reshuffled potential was a single, long helix. In another control where pairwise atomic potential was assigned a nonspecific value of -1 between any pair of atoms, we observed collapsed states for the same protein 1ENH with little secondary structure despite the fact that hydrogen bonding had the same strength as in full potential (data not shown). These control simulations indicate that it is likely that all three energy terms are crucial, and work in concert to give rise to the correctly folded structure.

### Why This Works

What is the reason that such a simple transferable energy function was able to fold many proteins at atomic resolution– a task of formidable complexity? Studies of simple models showed that kinetics and thermodynamics in protein folding are closely connected– protein-like sequences that are properly designed, i.e. that have large energy gap between native state and plethora of structurally dissimilar from the native state misfolds are able to fold quickly to their native states (Gutin et al., 1996; Shakhnovich, 1994; Shakhnovich, 2006). The relatively simple potential function presented here is effective in providing large gap between the native state and misfolds, as can be seen in Figure 3. There are several reasons for that. The first and foremost key physical idea behind the derivation of the μ-potential is the premise that sequences of native proteins were selected by nature to make their native structures separated by large energy gap from misfolds. For that reason, the μ-potential is perhaps one of the most efficient in identifying the set of unique interactions characteristic of native states of proteins. This assertion was confirmed by comparing various potentials for real proteins (Chen and Shakhnovich, 2005). Further analysis of the μ-potential derivation used the approach (Mirny and Shakhnovich, 1996) where a toy database of lattice proteins was designed and the μ-potential prescription was used to derive potentials from the database and compare them with "true" input potentials. The results (Deeds, Zeldovich, ES, unpublished) showed that the μ-potential prescription is an accurate procedure to recover potentials from the database of designed model proteins. Second, the energy gap is determined not only by the energy of the native state but also by the lowest energy states for misfolded decoys. The latter factor is affected by chain flexibility, i.e. number of allowed conformations per amino acid residue (Shakhnovich, 1994; Shakhnovich, 1998; Shakhnovich, 2006). The accurate hydrogen-bonding term as well as the knowledge-based local term significantly limit chain flexibility by energetically penalizing many chain conformations and therefore limiting the conformational space of decoys. This factor contributes towards increasing the energy gap making the complete potential sufficiently efficient in quickly guiding folding to the native conformation.

**Implications for protein folding.** In this work we presented a relatively simple energy function and a dynamic simulation that does not make any a'priori assumption about protein folding mechanism. The model is able to achieve near-native conformation starting from random coil conformations in a single run without refinement or output filtering or resolution change on the run. The native state prediction is identified as lowest energy conformation found in simulations. Perhaps the most important result of the present study is the finding that a transferable potential of a simple *form* – consisting of a pairwise atom-atom interactions taken in simplest, contact, form, accurate directional hydrogen bond and local stiffness energy is sufficient to fold proteins at least to moderate resolution and in some cases to high resolution structure..

Attempts to pinpoint dominant interactions that govern protein folding have been made in the literature (Dill, 1990). Several suggestions were presented including hydrophobic interactions as dominant force (Dill, 1990), packing interactions (Richards and Lim, 1993) as well as combination thereof and other possibilities. Understanding which interactions determine specificity of the native fold (Behe et al., 1991) has important implications for protein folding and structure predictions potentially providing a guidance into the necessary level of atomic and energetic detail that may be needed to predict protein structure. Indeed the ''dominant hydrophobic'' and ''dominant packing'' paradigms are two extremes, the former suggesting that native structure is encoded in a very simple coarse-grained hydrophobicity pattern of a sequence while the latter suggests that detailed shape complementarity of all amino acids forming the core of a protein is essential to determine protein structure. Our results as well as other analyses (Bradley et al., 2005; Zhang et al., 2005) suggest that contributions of multitude of interactions of comparable strengths rather than a single simple dominant force may be responsible for forming protein native structure. Indeed we found that sufficient diversity of atomic interactions (manifest in multitude of atom types, see Methods for details) along with accurate and explicit hydrogen bonding are necessary to fold proteins to near-native conformations. This is an important conclusion which is consistent with basic theoretical understanding that diversity of intramolecular and solvent interactions is necessary to provide energy gaps that enable a polypeptide chain to be stable in native conformation (Shakhnovich, 2006). On the other hand, approaches that use potential functions based on a small number of assumed dominant forces (Srinivasan et al., 2004) have so far showed only limited success in reproducing native like folds of proteins. It is also important to mention that some approaches to protein structure prediction (Srinivasan et al., 2004) are based on assumed – hierarchical – folding mechanism whereby secondary structure elements form first followed by their coalescence into tertiary structure. However, recent experimental and theoretical studies of protein folding kinetics suggest that this picture, while plausible, has a limited applicability to folding of many real proteins (Fersht and Daggett, 2002; Hubner et al., 2006; Meisner and Sosnick, 2004).

## Conclusions

The potential presented in this work appears to be able to fold small proteins at moderate resolution and in some cases (e.g. 1ENH and 1GJS) at high resolution. The robust results of this approach suggest that the basic physics such as specific compaction of hydrophobic residues provided by the μ-potential, inclusion of hydrogen-bonding specific to α- and β-conformations, highly local effects captured in a sequence dependent potential, and an all-atom model, are essential factors to predict the structure of small compact proteins with various topologies.

The energy function presented here is still phenomenological in the sense that it is presented in a simplified *form*. The trade-off for that is the need to introduce large number of atom types in order to maintain the necessary diversity of interactions to provide sufficient energy gap, as explained above. However this energy function appears to be simple enough to make *ab initio simulation* of complete folding process – without any assumed mechanism or local or global templates – possible In this sense our approach is a truly an ab initio one. An accurate and more physically fundamental energy function with less diverse set of atoms may emerge in the future. However the *form* of such energy function –most likely including explicit solvent, polarisable distant-dependent force-field, maybe even quantum effects - is bound to be too complicated to make ab initio simulation of complete folding out of reach with available computational resources for some time to come.

The sampling of larger proteins, without substantially increasing the computational requirements, might be accomplished by extending the knowledge-based move-set (Chen et al., 2006) to pairs or triplets of $\phi/\psi$ angles. The knowledge-based move-set for single $\phi/\psi$ angles makes the search more efficient (Chen et al., 2006). Therefore, we find it reasonable that sampling the most frequently observed pairs and triplets of $\phi/\psi$ angles would lead to an even more efficient search of the ground state, which will be done in a future study.

The successful *ab initio* all-atom folding reported in this work opens an exciting opportunity to address key questions in protein folding dynamics (Hubner et al., 2006). It brings us closer to a comprehensive *ab initio* solution of the protein folding problem, including atomic resolution analysis of the folding mechanism– from random coil to the fully folded native state.

## Experimental Procedures

### The Energy Function

The energy function used in atomic simulations has the form:

$$U = E_{\text{con}} + a \times E_{\text{trp}} + b \times E_{\text{hb}} \quad (1)$$

where $E_{\text{con}}$ is the pairwise atom-atom contact potential, $E_{\text{hb}}$ is the hydrogen-bonding potential, and $E_{\text{trp}}$ is the sequence-dependent local conformational potential based on the statistics of sequential amino acid triplets.

### The Contact Energy

The contact energy, so-called the μ-potential for multi-proteins is defined as

$$E_{\text{con}} = \sum_{i<j} E_{\text{A}_i\text{A}_j}, \quad E_{\text{AB}} = \frac{-\mu N_{\text{AB}} + (1-\mu)\tilde{N}_{\text{AB}}}{\mu N_{\text{AB}} + (1-\mu)\tilde{N}_{\text{AB}}}, \quad (2)$$

where $A_i$ is the atom type of an atom $i$ and $N_{\text{AB}}$ and $\tilde{N}_{\text{AB}}$ are the total number of contacts and the total number of pairs not in contact in the data base consisting of 6260 proteins, respectively, i.e. $N_{AB} = \sum_{i=1}^{6260} N_{AB}^{i}$ and $\tilde{N}_{AB} = \sum_{i=1}^{6260} \tilde{N}_{AB}^{i}$ . Two atoms A and B are defined to be in contact if the distance between them is less than $\lambda(r_{\text{A}} + r_{\text{B}})$, where $r_{\text{A}}$ and $r_{\text{B}}$ are their respective van der Waals radii. We took $\lambda = 1.8$ as in our previous work (Shimada et al., 2001). Compared to recently developed atom pair potentials such as DFIRE (Zhou and Zhou, 2002), we use a single-distance bin which saves a lot of computing time. To prevent the significant overlap between atoms, atomic hard sphere radii were taken to be 0.75 of their van der Waals sizes (Shimada et al., 2001). An atom typing scheme in which each side-chain atom of each of the 20 amino acids is assigned a separate type along with four backbone atoms is used, resulting in total of 84 atom types (Kussell et al., 2002). The multi-protein formulation of the contact energy, the so-called the μ-potential was shown to perform better than a quasi-chemical potential in all-atom threading and decoy discrimination tests (Chen and Shakhnovich, 2005). The value of $\mu = 0.995$ was chosen to make the net interaction zero.

**The Triplet Local Conformational Energy**

The triplet energy term uses information from short fragments of length three and is amino acid specific. A triplet consisting of $A_i$, $A_{i+1}$, and $A_{i+2}$ is shown in Figure 5, where $A_i$ is the amino acid type of a residue *i*. The bold letter $\mathbf{b}_{A_i}$ is a vector bisecting two vectors ($\mathbf{C}^{\alpha}_{A_i}\mathbf{N}_{A_i}$ and $\mathbf{C}^{\alpha}_{A_i}\mathbf{C}_{A_i}$) and $\mathbf{P}_{A_i}$ represents a vector in a plane defined by three backbone atoms ($\mathbf{N}_{A_i}$, $\mathbf{C}^{\alpha}_{A_i}$, and $\mathbf{C}_{A_i}$) for residue *i*. Four variables in this energy function are $\phi_{A_{i+1}}$, $\psi_{A_{i+1}}$, the angle between $\mathbf{b}_{A_i}$ and $\mathbf{b}_{A_{i+2}}$, and the angle between $\mathbf{P}_{A_i}$ and $\mathbf{P}_{A_{i+2}}$. The width of bins for $\phi_{A_{i+1}}$ and $\psi_{A_{i+1}}$ was 60° and the value of 30° is used for the other two angles. The triplet energy was obtained from the database by

$$E_{\mathrm{trp}} = \sum_i E_{A_i A_{i+1} A_{i+2}}, \quad E_{A_i A_{i+1} A_{i+2}} = \frac{-\mu N_j + (1-\mu)\widetilde{N}_j}{\mu N_j + (1-\mu)\widetilde{N}_j}, \tag{3}$$

where $N_j$ and $\widetilde{N}_j$ are the number of observations in the *j*-th bin and total number of observances subtracted by $N_j$ for a triplet consisting of $A_i$, $A_{i+1}$, and $A_{i+2}$, respectively. The value of $\mu = 0.991$ was chosen to make the net interaction zero. There are about 1.5 million triplets in the potential database consisting of 6260 sequences (See below). The total no. of bins (≈ 10 million) is even bigger than this. However, note that the four angle bins are not independent from each other and also most of bins are inaccessible due to steric overlaps of side-chain atoms, i.e. only a part of the φ/ψ space is populated in the Ramachandran plot. This significantly reduces the actual no. of bins for this potential, which enables us to use this potential without an undersampling issue. For example, the most populated bin corresponding to the helix conformation for a triplet of "AAA" has 651 observations. A notable difference between our sequence-dependent potential and ROSETTA's approach (Simons et al., 2001) is that we do not restrict the conformational space of the fragments to the conformations observed in real fragments, as ROSETTA does. As such, triplet configurations are recorded and sampled over all possible conformations making it possible to carry out fully *ab initio* simulations rather than fragment buildup procedure.

**The Hydrogen-bonding Energy**

The hydrogen-bonding of a donor-acceptor pair is properly taken into account by adapting the residue-based hydrogen-bonding model of Zhang and Skolnick (Zhang and Skolnick, 2004; Zhang and Skolnick, 2005) to the detailed all-atom model. Figure 6 shows a hydrogen-bonded pair of $O_i$ and $H_j$ and its corresponding geometry. If the distance between oxygen and hydrogen atoms is less than 2.5 Å, then the following six variables are used to determine the hydrogen-bonding energy.

1. The angle between $\mathbf{b}_i$ and $\mathbf{b}_{j-1}$.
2. The angle between $\mathbf{b}_{(i)(i+1)}$ and $\mathbf{b}_{(j-1)(j)}$.
3. The angle between $\mathbf{b}_{i+1}$ and $\mathbf{b}_j$.
4. The angle between $\mathbf{P}_i$ and $\mathbf{P}_{j-1}$.
5. The angle between $\mathbf{P}_{(i)(i+1)}$ and $\mathbf{P}_{(j-1)(j)}$.
6. The angle between $\mathbf{P}_{i+1}$ and $\mathbf{P}_j$.

The width of bins for these variables was 20°. The hydrogen-bonding energy was obtained by the quasi-chemical approximation from the database. In addition, we introduced a weighting factor $c$ as follows: if the pair of donor and acceptor residues is more than 4 residues apart (which corresponds to hydrogen-bonding in a β-sheet), the hydrogen-bond energy is multiplied by a factor $c > 1$. The reason for the use of this weight factor is to ensure that hydrogen-bonds in α-helices and β-strands are equally weighted, whereas both configurational entropy and the triplet energy obtained from ϕ/ψ probabilities favor the α-helix over the β-sheet (Shortle, 2003). The hydrogen-bonding term was weakly modulated by secondary structure prediction results from PSIPRED (McGuffin et al., 2000) in the following way. A residue predicted as "helix" will have zero hydrogen-bonding energy to residues which are located more than four residues away along the sequence. A residue predicted as "sheet" will have zero hydrogen-bonding energy to a residue in helical conformation that is exactly four residues away along the sequence. Only predictions from PSIPRED with confidence levels greater than three were used. The main reason that the PSIPRED input is used here is to help the chain to avoid transient helical conformations for residues that are confidently predicted to be in extended β-conformations as well as to avoid transient β-conformations for residues that are confidently predicted as helical. The PSIPRED input has mostly kinetic effect by reducing local traps due to transient incorrect local secondary structure formation by avoiding occasional sheet conformations in a helical stretch and vice versa. This provides an important kinetic advantage given the limited sampling available. While PSIPRED input may bias selection between β and α conformations for residues that are part of secondary structure elements, it does not bias selection of conformation of an amino acid residue between coil and a secondary structured one. Indeed, we note that a residue predicted as "helix" or "sheet" can convert to "coil" with no penalty from PSIPRED input, and that a residue predicted as "coil" also can convert into α- or β-conformation without PSIPRED-related penalty. Ultimately energetic balance of hydrogen-bonding, local interaction, and contact potential contribution decides whether a residue should be in a coiled conformation or belong to secondary structure. Consequently, the secondary structures in this simulation are quite different from those of PSIPRED, which means that the secondary structures assignments in our simulation are determined by the energy function, not by PSIPRED (Figure 7).

**The Potential Database**

The potential parameters in the three energy terms were obtained from the PISCES database (Wang and Dunbrack, 2003) collected from 6260 sequences determined by X-ray with the resolution cutoff of 3.0 Å that contains none of the eighteen proteins used as the training and testing sets. To check the effects of the homologous sequences in the database on the results, we rederived new sets of potential parameters after removing 76 sequences that show sequence identity greater than 30 to the proteins studied in this work (about 5 homologous sequences out of 6260 sequences for each protein; less than 0.1 %). We confirmed that energies of the native structures and *Z*-scores from the decoy sets are almost identical (data not shown). We also performed simulations starting from the native with the new potential and got comparable results to those with the original potential (data not shown).

**Move Sets**

The degrees of freedom in this simulation are the $\psi$ and $\chi$'s of all residues and $\phi$ except in proline. Three kinds of moves are used: backbone moves ($\phi$ and $\psi$), side-chain moves ($\chi$'s) and "knowledge-based" moves (Chen et al., 2006). A backbone move is either global or local with equal probability. A global move for the backbone consists of rotating the dihedral angle ($\phi$ or $\psi$) of a randomly selected residue (Shimada et al., 2001). A local move has an effect on four successive residues within a window, with atoms outside this window unchanged (Coutsias et al., 2004). The step sizes of the global and local moves for the backbone are drawn from a Gaussian distribution with zero mean and standard deviation of 2 and 60 degrees, respectively. A side-chain move consists of rotating all $\chi$ angles in a randomly selected non-proline residue (Shimada et al., 2001). The step size of the side-chain rotation is drawn from a Gaussian distribution with zero mean and standard deviation of 10 degrees. A knowledge-based dihedral move (Chen et al., 2006), similar to a global backbone move, was also included. The knowledge-based moves were formulated in the following way: backbone dihedral angles for each of the twenty residues were recorded for hundred randomly selected proteins from the same database used for deriving the potential energy function. For each residue, the observed dihedral angles were clustered by K-means (Jain and Dubes, 1988) in $\phi/\psi$ space down to thirty representative points. These thirty representative points are the clustered $\phi/\psi$ angles for the particular residue. A knowledge-based move of a residue during simulation entails setting the dihedral angles of the residue randomly to one of the clustered $\phi/\psi$ angles. Comparison of REMC simulations with and without the knowledge-based move set shows that simulations using the knowledge-based move set, in most cases, find states with much lower energy (data not shown). Additionally, simulations without the knowledge-based move set were not as effective at forming proper secondary structures. However, inclusion of the knowledge-based move set violates detailed balance. While this is not a serious limitation for the purposes of the present study that seeks conformations with minimal possible energy for structure prediction purposes, simulations utilizing this move cannot be used to assess thermodynamic quantities unless the move set is further modified to restore detailed balance, which should be solved in the future work.

**Initial Configuration**

The initial configuration is prepared as follows. A long, single helix structure of the sequence is generated using SWISS-PDB VIEWER (Guex and Peitsch, 1997). This structure has many severe clashes and needs to be relaxed before the replica exchange Monte Carlo (REMC) simulation begins. This is done by MC simulations at very high temperature ($t = 1000$), which result in a random coil. The structure is then energy minimized by CHARMM (Brooks et al., 1983) to correct variables which are not degrees of freedom in the REMC simulation (bond lengths, bond angles, ω angles, and the ϕ angle in proline). For each protein, three independent initial configurations were prepared as described above and five REMC simulation runs were done for a given initial configuration, resulting in fifteen REMC simulations.

**REMC Simulation**

We set up REMC simulations (Gront et al., 2000; Swendsen and Wang, 1986) with 30 replicas at different temperatures, ranging from 0.15 to 1.50. The trajectories at the lowest temperature ($t = 0.150$) were analyzed for structure prediction. Each REMC simulation takes approximately 10 hours on 30 2.4 GHz Xeon CPUs, which results in about 190 CPU days for each protein.

**Training and Testing Sets**

Three adjustable parameters in the energy function (*a*, *b*, and *c*) were roughly optimized for the training set by the REMC simulations. Parameters were optimized by minimizing RMSD values of the minimum energy structures in the training set and ensuring consistency of secondary structure elements between them. Five proteins were used for the training set: PDB ID codes 1E0L, 1I6C (β proteins), 1E0G (α+β protein), 1BDD, and 1ENH (α proteins). The optimized values were $a = b = c = 3.0$. REMC simulations for the testing set used the same energy function. Thirteen proteins were used for the testing set: PDB ID codes 1K9R, 1SHFA, 1CSP (β protein), 1IGD, 1TIF, 1CLB (α/β and α+β proteins), 1GAB, 1BA5, 1GJS, 1GUU, 1LQ7, 1AIL, and 1LFB (α proteins). The number of residues in the training and testing sets ranges from 24 for 1K9R to 77 for 1LFB (Table 1).

## Acknowledgements

We thank Y. Zhang, H. Kaya, C. Seok, and M. Heo for helpful discussions.  This research was supported in by grants from National Institutes of Health.

## Figure Legends

Figure 1. Successful folding to native state of proteins from the training set

The lowest-energy structure from the trajectory at the lowest temperature is compared with the experimental structure. Red to blue runs from the N to the C terminus. (A) 1I6C, composed of 3 β-strands, was folded to an RMSD of 2.72 Å. (B) 1E0G, an α+β protein, was folded to an RMSD of 4.15 Å. Black arrows indicate a small portion of the helix was not properly formed *in silico*. (C) 1ENH, a helical protein, was folded to an RMSD of 2.88 Å. Structures are displayed by using VMD (Humphrey et al., 1996).

Figure 2. Successful folding to native state of proteins from the test set

(A) The relatively complex topology of 1IGD was reproduced in simulation with an RMSD of 3.92 Å. (B) All secondary structure elements of 1CLB except for two short strands are correctly predicted with an RMSD of 5.41 Å. Black arrows indicate missing strands. (C/D) Predicted structures of 1GJS and 1LQ7 have very good agreement with their experimental structures with RMSD values of 2.26 Å, and 3.86 Å, respectively.

Figure 3. The energy landscape for the proteins in *ab initio* REMC simulations

Blue: starting from the native, red: from a random coil. (A/B) The energy shows good correlation with RMSD for 1IGD and 1GJS, respectively. (C) A mirror image topology was observed at 8.0 Å for 1LQ7, but it has higher energy than the native-like structure. (D) Simulations from the native state show that conformations near the native state form a well-defined minimum at low RMSD compared to those from a random coil for 1CSP.

Figure 4. Examples of proteins imperfectly predicted in the test set

(A) The predicted lowest energy structure for 1BA5 is the mirror-image topology. The RMSD is 8.83 Å. (B) The C-terminal part of the β-sheet is not correctly formed in 1SHFA. (C) Two sheets in the N terminus have different alignment from the native for 1CSP. (D) The red helix is not properly formed in the predicted structure for 1AIL. The RMSD is 7.48 Å.

Figure 5. A schematic diagram for the triplet energy term

Figure 6. Geometrical considerations for the hydrogen-bonding energy for an interacting pair of $O_i$ and $H_j$

Figure 7. Comparison of the secondary structures obtained from *ab initio* simulations with those predicted by PSIPRED

(A) Secondary structures of the all the residues for 1I6C, 1BDD, 1ENH, and 1GAB. Note that the secondary structures in *ab initio* simulations are different from the PSIPRED results. (B) The Q3 values, the fraction of residues predicted to be in the correct conformational state (helix, strand or coil), for all of 18 proteins. The secondary structures of simulated and native PDB files were assigned by STRIDE (Frishman and Argos, 1995).

Table 1 Summary of simulation results for the training and testing sets

| Protein | $N^a$ | Class | $R_{min}^{b,c}$ (Å) | $S_{R_{min}}^{d,c}$ | $R_{E_{min}}^{e,c}$ (Å) | $S_{E_{min}}^{f,c}$ | $E_{min}^{g,c}$ | $R_{E_{min}}^{native\ e,h}$ (Å) | $E_{min}^{native\ g,h}$ |
|---|---|---|---|---|---|---|---|---|---|
| 1E0L | 25 | β | 1.15 | 0.92 | 4.19 | 0.44 | -116 | 3.13 | -100 |
| 1I6C | 25 | β | 2.17 | 0.73 | 2.72 | 0.65 | -118 | 5.50 | -126 |
| 1BDD | 47 | α | 2.33 | 0.70 | 6.41 | 0.40 | -176 | 3.79 | -173 |
| 1E0G | 48 | α+β | 3.44 | 0.49 | 4.15 | 0.46 | -192 | 3.11 | -195 |
| 1ENH | 48 | α | 1.92 | 0.80 | 2.88 | 0.60 | -179 | 2.12 | -202 |
| 1K9R | 24 | β | 2.40 | 0.73 | 4.10 | 0.51 | -117 | 3.34 | -100 |
| 1GAB | 46 | α | 3.32 | 0.61 | 4.97 | 0.51 | -157 | 1.94 | -167 |
| 1BA5 | 46 | α | 5.35 | 0.38 | 8.83 | 0.31 | -197 | 7.97 | -204 |
| 1GJS | 49 | α | 1.94 | 0.78 | 2.26 | 0.72 | -181 | 2.15 | -179 |
| 1GUU | 50 | α | 4.17 | 0.25 | 4.70 | 0.23 | -185 | 4.19 | -182 |
| 1IGD | 57 | α/β | 3.79 | 0.47 | 3.92 | 0.47 | -268 | 1.64 | -285 |
| 1SHFA | 59 | β | 5.86 | 0.30 | 6.08 | 0.28 | -250 | 1.81 | -273 |
| 1TIF | 59 | α+β | 4.67 | 0.17 | 5.64 | 0.18 | -237 | 2.39 | -240 |
| 1LQ7 | 67 | α | 3.48 | 0.29 | 3.86 | 0.27 | -241 | 1.96 | -266 |
| 1CSP | 67 | β | 6.97 | 0.28 | 8.23 | 0.30 | -270 | 3.71 | -323 |
| 1AIL | 70 | α | 6.49 | 0.30 | 7.48 | 0.31 | -264 | 2.23 | -272 |
| 1CLB | 75 | α+β | 4.64 | 0.38 | 5.41 | 0.28 | -325 | 3.13 | -326 |
| 1LFB | 77 | α | 5.47 | 0.25 | 6.37 | 0.25 | -285 | 3.09 | -293 |

[a]Number of residues in the protein.

[b]Minimum RMSD seen in the simulation.

[c]Obtained from fifteen trajectories at the lowest temperature ($t = 0.150$) starting from random coil structures.

[d]MaxSub score at the RMSD minimum (cutoff = 3.5 Å).

[e]RMSD value at the energy minimum.

[f]MaxSub score at the energy minimum (cutoff = 3.5 Å).

[g]Minimum of energy value.

[h]Obtained from the trajectory at the lowest temperature ($t = 0.150$) starting from the native structure as a control.

Figure 1-Yang etc

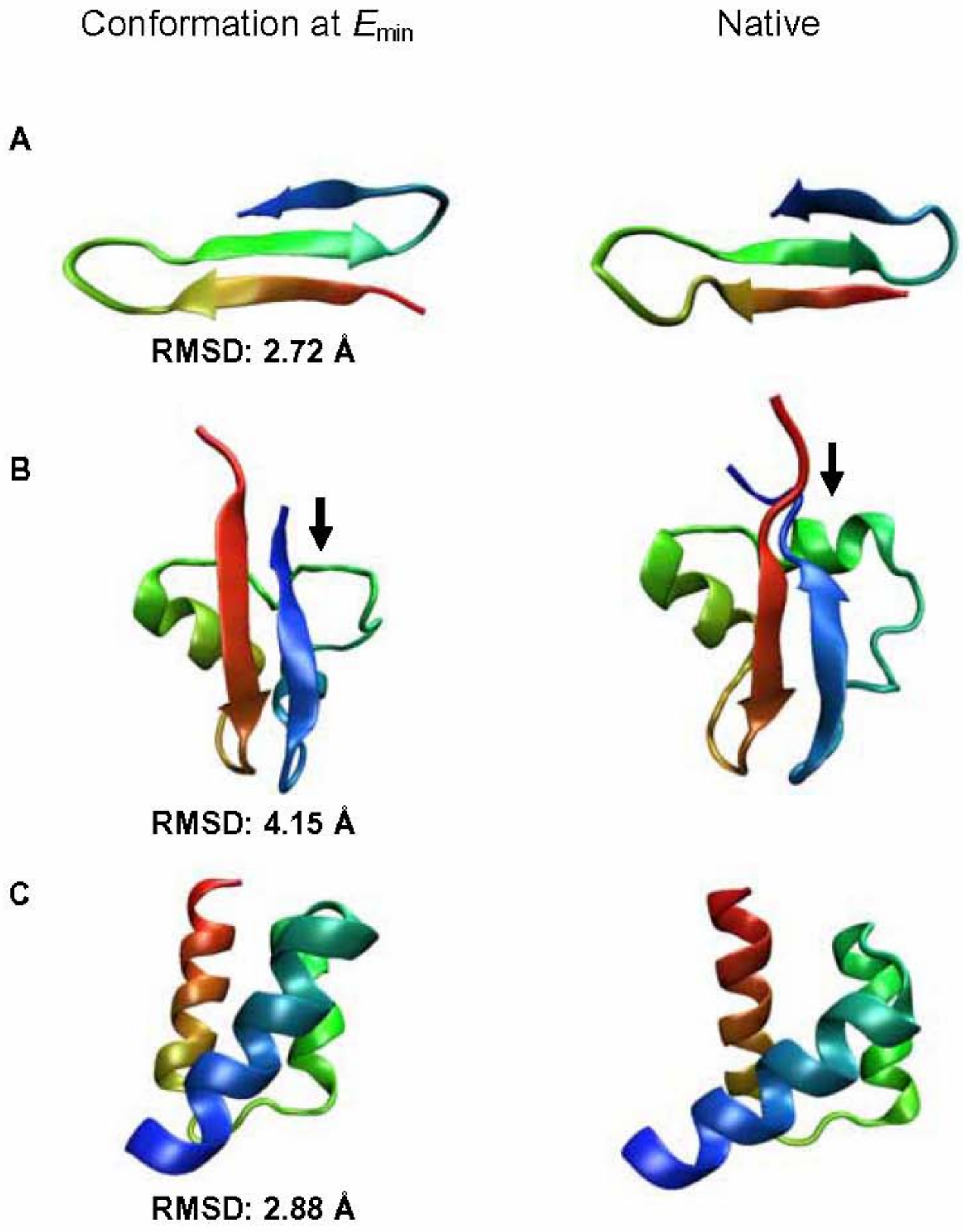

1

Figure 2-Yang etc

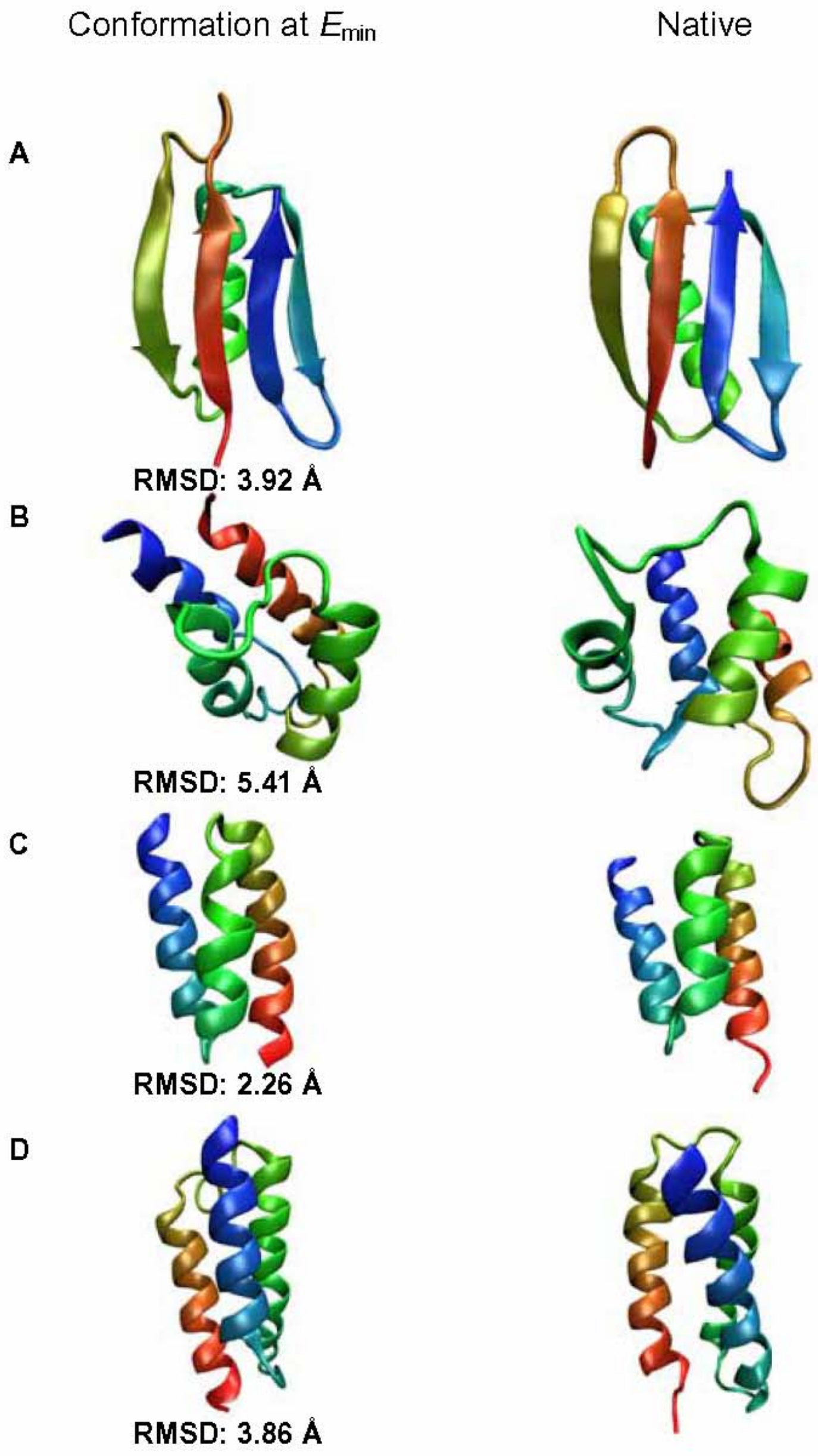

2

**Figure 3-Yang etc**

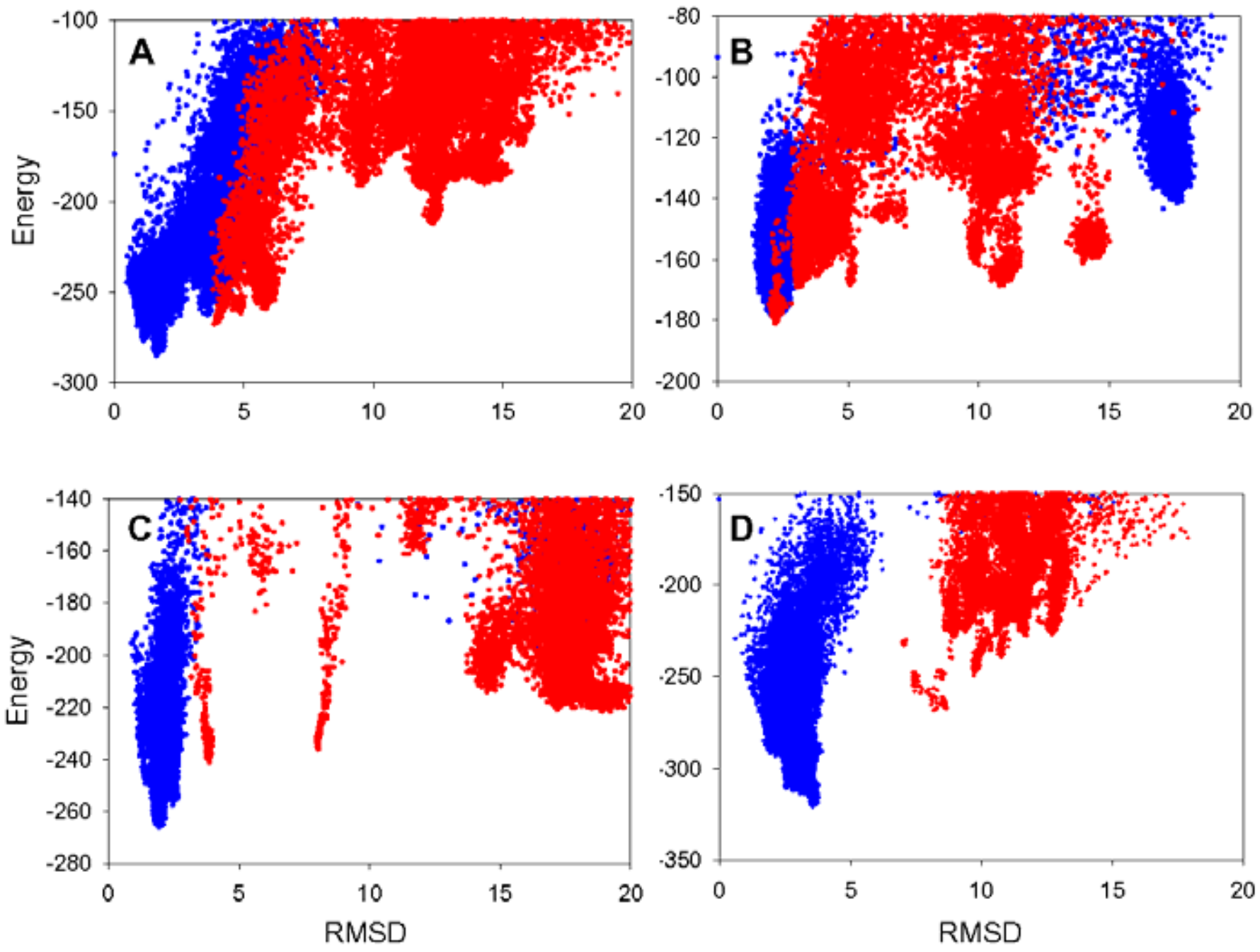

3

Figure 4-Yang etc

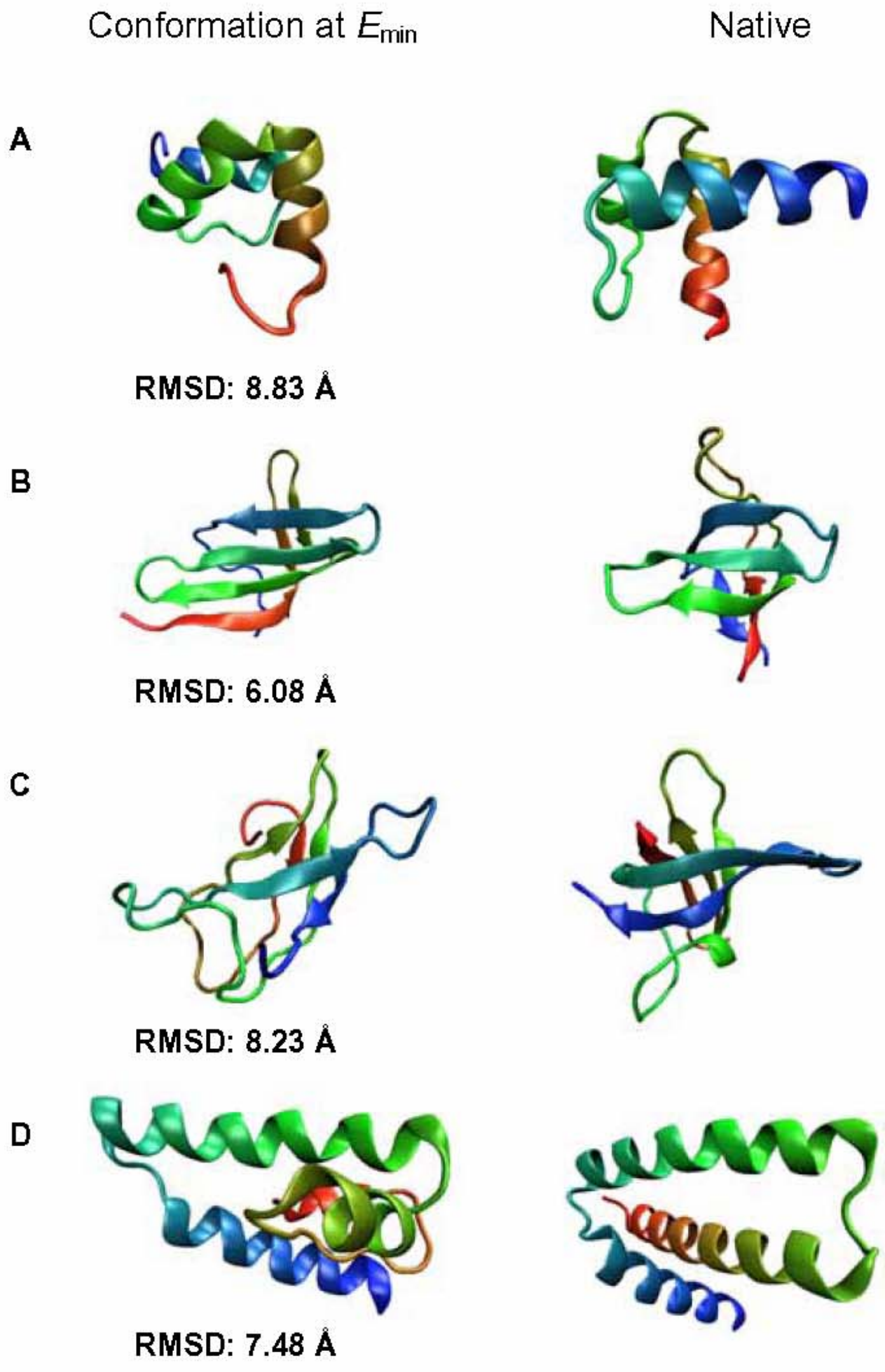

4

**Figure 5-Yang etc**

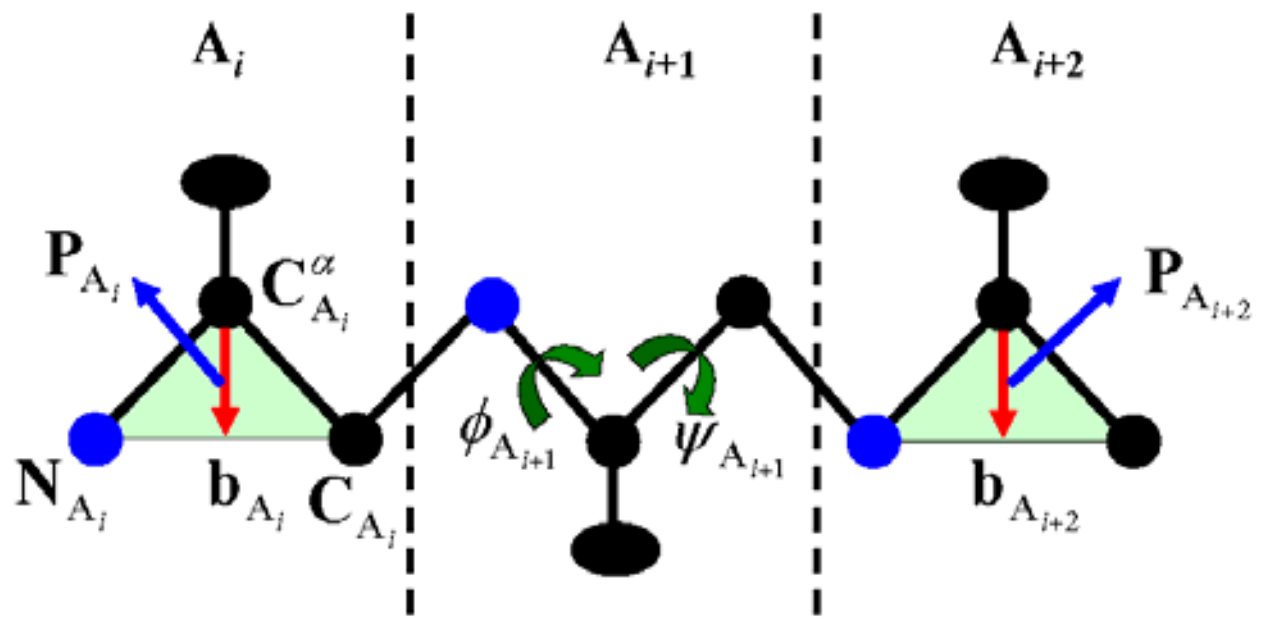

5

**Figure 6-Yang etc**

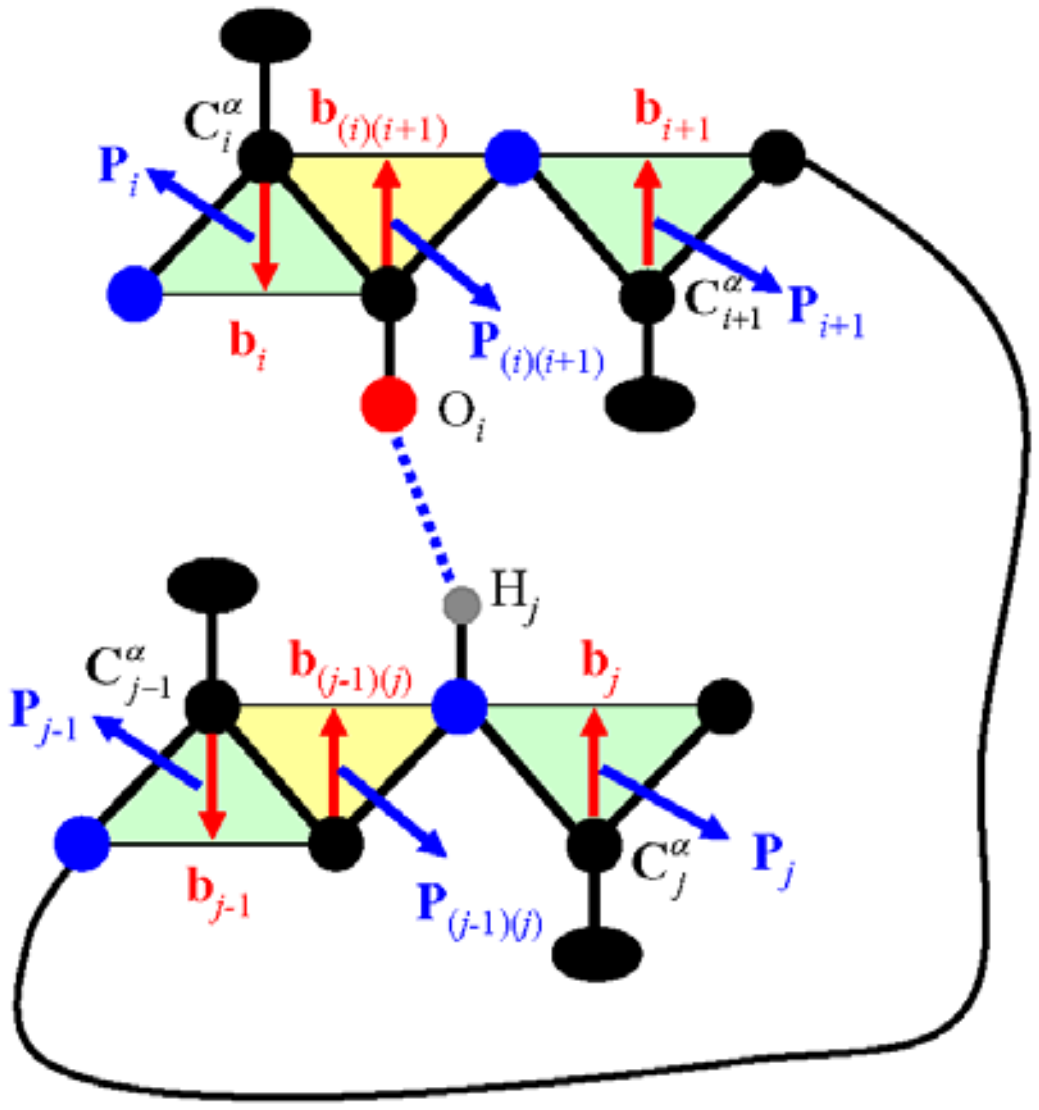

6

**Figure 7-Yang etc**

A

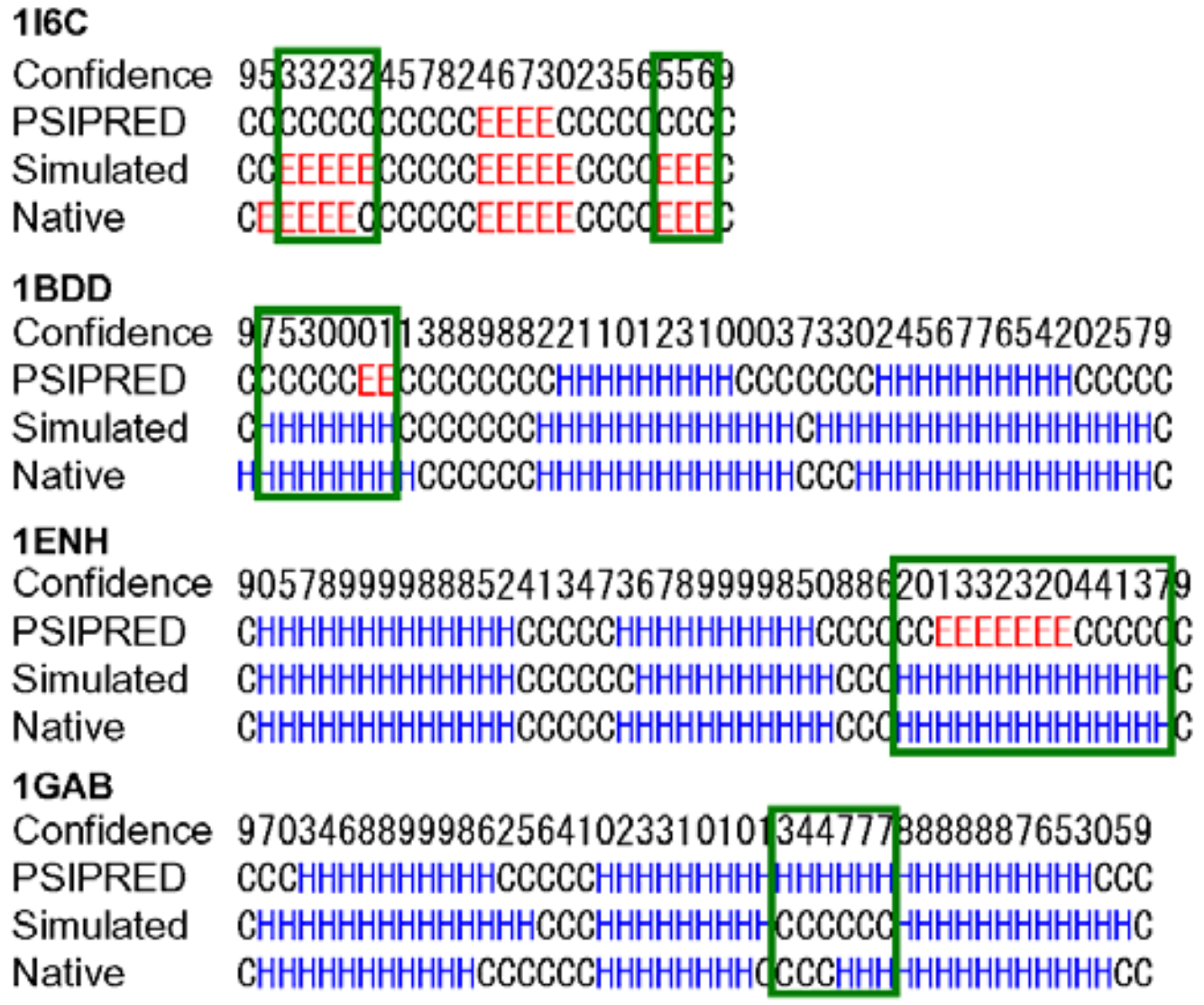

B

| Protein | *N* | Q3 | | Protein | *N* | Q3 | |
|---|---|---|---|---|---|---|---|
| | | PSIPRED | Simul. | | | PSIPRED | Simul. |
| 1E0L | 25 | 48 | 88 | 1GUU | 50 | 74 | 86 |
| 1I6C | 25 | 64 | 92 | 1IGD | 57 | 79 | 89 |
| 1BDD | 47 | 62 | 91 | 1SHFA | 59 | 78 | 75 |
| 1E0G | 48 | 85 | 62 | 1TIF | 59 | 86 | 75 |
| 1ENH | 48 | 69 | 98 | 1LQ7 | 67 | 94 | 81 |
| 1K9R | 24 | 50 | 79 | 1CSP | 67 | 85 | 54 |
| 1GAB | 46 | 83 | 83 | 1AIL | 70 | 79 | 81 |
| 1BA5 | 46 | 91 | 79 | 1CLB | 75 | 88 | 83 |
| 1GJS | 49 | 88 | 90 | 1LFB | 77 | 83 | 74 |

7